\documentclass[aoas,MSNbibl,nameyear,seceqn,dvips]{arximspdf}
\usepackage{graphicx}

\doi{10.1214/12-AOAS573} 
\volume{7}
\issue{1}
\pubyear{2013}
\firstpage{201}
\lastpage{225}

\makeatletter

\newcommand{\rright}{\right}
\newcommand{\lleft}{\left}
\newtheorem{theorem}{Theorem}

\newproclaim{remark}{Remark}

\newtheorem{corollary}{Corollary}
\newproclaim{example}{Example}
\newproclaim{definition}{Definition}
\def\hat{\widehat}
\def\tilde{\widetilde}
\def\conD{\stackrel{\mathcal{L}} \to}
\def\conP{\stackrel{\mathcal{P}} \to}
\def\ol{\overline}
\def\bD{\mathbf{D}}
\def\ba{\mathbf{a}}
\def\bb{\mathbf{b}}
\def\bU{\mathbf{U}}
\def\bg{\mathbf{g}}
\def\bx{\mathbf{x}}
\def\bv{\mathbf{v}}
\def\bzero{\mathbf{0}}
\def\eps{\varepsilon}
\def\blambda{\bolds{\lambda}}

\def\MSE{\operatorname{MSE}}
\def\RA{\operatorname{RA}}
\def\FA{\operatorname{FA}}
\def\FDR{\operatorname{FDR}}
\def\rank{\operatorname{rank}}
\def\chiK{\mathbb{K}}
\def\ID{ \mathrm{I}}
\def\calV{\mathcal{V}}
\def\mathN{{\mathcal{N}}}
\def\mathS{{\mathcal{S}}}
\def\SNR{\operatorname{SNR}}
\def\rankA{\mu}
\def\trace{\operatorname{trace}}
\def\btheta{\bolds{\theta}}
\def\diag{\operatorname{diag}}
\makeatother

\begin{document}
\begin{frontmatter}

\title{Local tests for identifying anisotropic diffusion areas in human brain with DTI}
\runtitle{Local tests for anisotropic areas}

\begin{aug}
\author[A]{\fnms{Tao} \snm{Yu}\corref{}\ead[label=e1]{stayt@nus.edu.sg}\thanksref{t1}},
\author[B]{\fnms{Chunming} \snm{Zhang}\ead[label=e2]{cmzhang@stat.wisc.edu}\thanksref{t2}},
\author[C]{\fnms{Andrew~L.}~\snm{Alexander}\ead[label=e3]{alalexander2@wisc.edu}\thanksref{t3}}
\and
\author[C]{\fnms{Richard J.} \snm{Davidson}\ead[label=e4]{rjdavids@wisc.edu}\thanksref{t3}}
\thankstext{t1}{Supported in part by NUS Grant R-155-000-100-133.}
\thankstext{t2}{Supported in part by NSF Grant DMS-11-06586; Wisconsin Alumni
Research Foundation.}
\thankstext{t3}{Supported in part by National Institute of Mental
Health Grants R01-MH43454 and P50-MH069315 to RJD and by Grant P30
HD003352 to the Waisman Center (PI: M. Seltzer).}

\runauthor{Yu, Zhang, Alexander and Davidson}
\affiliation{National University of Singapore, University of
Wisconsin-Madison, University of Wisconsin-Madison and University of
Wisconsin-Madison}
\address[A]{T. Yu\\
Department of Statistics \\
\quad and Applied Probability\\
National University of Singapore\\
Singapore 117546\\
\printead{e1}}
\address[B]{C. Zhang \\
Department of Statistics \\
University of Wisconsin-Madison\\
Madison, Wisconsin 53706\\
USA\\
\printead{e2}}

\address[C]{A. L. Alexander\\
R. J. Davidson \\
Waisman Center\\
University of Wisconsin-Madison\\
Madison, Wisconsin 53705\\
USA\\
\printead{e3}\\
\phantom{E-mail:\ }\printead*{e4}}

\end{aug}

\received{\smonth{9} \syear{2011}}
\revised{\smonth{4} \syear{2012}}

%
\begin{abstract}
Diffusion tensor imaging (DTI) plays a key role in analyzing the physical
structures of biological tissues, particularly in reconstructing fiber tracts
of the human brain in vivo. On the one hand, eigenvalues of diffusion
tensors (DTs) estimated from diffusion weighted imaging (DWI) data
usually contain systematic bias, which subsequently biases the
diffusivity measurements popularly adopted in fiber tracking
algorithms. On the other hand, correctly accounting for the spatial
information is important in the construction of these diffusivity
measurements since the fiber tracts are typically spatially structured.
This paper aims to establish test-based
approaches to identify anisotropic water diffusion areas in the human
brain. These areas
in turn indicate the areas passed by fiber tracts. Our proposed test
statistic not only takes into account
the bias components in eigenvalue estimates, but also incorporates the
spatial information of neighboring voxels.
Under mild regularity conditions, we demonstrate that the proposed test
statistic asymptotically
follows a $\chi^2$ distribution under the null hypothesis. Simulation
and real DTI data examples are provided to illustrate
the efficacy of our proposed methods.
\end{abstract}

%

\end{frontmatter}

\section{Introduction} \label{sec-1}

\begin{figure}

\includegraphics{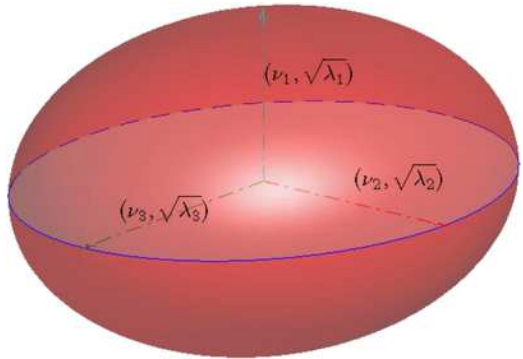}

\caption{Ellipsoid representation of a DT.}\label{Figure-Tensor-disp}
\end{figure}

Diffusion tensor imaging (DTI) has been widely used by
neuroscientists to reconstruct the pathways of white matter fibers
in human brain in vivo. DTI data are usually estimated from diffusion
weighted imaging (DWI) data acquired in magnetic resonance experiments
by some statistical model (see Section~\ref{sec-2-1}). A set of DTI
data is typically composed of diffusion tensors (DTs), each contained
in a corresponding voxel. Here, a voxel stands for a volume element in
a 3D imaging space. Each DT, denoted by $\bD$, can be represented by a
$3\times
3$ symmetric positive definite matrix, which together with its decomposed
eigenvalue--eigenvector pairs $\{(\lambda_{(k)}, \bv_{(k)})\dvtx  \lambda_{(3)} \ge\lambda_{(2)} \ge
\lambda_{(1)}, k=1,2,3\}$ geometrically characterizes the
degree and orientation of the water diffusion in that particular voxel.
More specifically, the eigenvectors and the square root of eigenvalues
of $\bD$, respectively, correspond to the orientations and the lengths of
axes in an ellipsoid representation (see Figure \ref
{Figure-Tensor-disp}). The distance between the center and any point on
the surface of the ellipsoid measures the rate of the water diffusion
along that particular orientation in the voxel.

%

One of the main themes in DTI research is to identify the anisotropic
water diffusion brain areas, which facilitates the downstream fiber
tracking process. There are two general strategies aiming to address
this problem. The first is to construct scalar measurements. The
anisotropic water diffusion areas are then identified based on
thresholding these scalar measurements. Reviews of this type of method can
be found in \citet{Moseley1990}, \citet{Douek1991}, \citet
{vanGelderen1994}, \citet{BasserPierpaoli1996},
\citet{JohansenBergBehrens2009}, among others.
Thresholding the fractional anisotropy (FA) and the relative anisotropy
(RA) [see Section~\ref{sec-2-2};
\citet{BasserPierpaoli1996}] has gained popularity and has been widely
adopted by neuroscientists
in the past decade. Nonetheless, FA and RA are essentially defined as
functions of the eigenvalues of the DT estimate in every brain voxel.
These eigenvalues typically carry systematic bias [see Section \ref
{sec-3}; \citet{PierpaoliBasser1996}, \citet{Zhu2007},
\citet{Jones2003},
\citet{LazarAlexander2003}], the magnitudes of which are sensitive to
the distribution of the noise carried in raw DWI data, and therefore
significantly affect the effectiveness and validity of FA (RA) based
methods in practice. The second is to classify the morphologies of DTs
via test-based approaches [\citet{Zhu2006}, \citet{Zhu2007}],
which not only quantify the degree of water
diffusivity in each voxel, but also lend theoretical supports to statistical
inference. However, to the best of our knowledge, all existing
approaches on this respect are single-voxel based. The validity and
performance of these methods rely essentially on the technical
requirement that the number of diffusion gradients in the DWI
experiment is large and ultimately diverging to infinity. Moreover, the
DTI data are typically spatially structured. Ignoring the spatial
information may diminish the effectiveness of the methods.\looseness=1

In this paper, we develop new test-based approaches to identify the
aniso\-tropic water diffusion brain areas, which are usually associated
with areas passed by fiber tracts. To this end, for each voxel, we
examine the testing problem ``\textit{all three eigenvalues are
equivalent}'' against ``\textit{at least two eigenvalues are
different}.'' The former corresponds to isotropically diffused DTs,
whereas the latter includes anisotropically diffused DTs with
morphologies of prolate, oblate and nondegenerate. Our proposed test
statistic accommodates the spatial information of the imaging space by
taking into account eigenvalues in neighboring voxels. Under mild
regularity conditions, we demonstrate that our proposed test statistic
asymptotically follows a $\chi^2$ distribution. Therefore, the
performance of our methods in the identification of fiber areas is not
affected by the bias components carried in the eigenvalue estimates. In
theory, one of the main technical requirements is the divergence of the
number of neighboring voxels involved in the construction of the
statistic. This differs from the divergence of the number of diffusion
gradients typically assumed by test-based approaches. Therefore, our
methods shed light on alternative ways of improving the identification
accuracy of anisotropic water diffusion areas. Furthermore, an adaptive
procedure to select varied neighborhoods is proposed to solidify the
performance of our proposed approaches when the acquired imaging data
have limited resolution. Simulation studies and real data examples are
provided to illustrate the efficacy of our proposed methods.

The rest of the paper is organized as follows. Section~\ref{sec-2}
introduces the background related to our study.
Section~\ref{sec-3} establishes a statistical model based
on eigenvalues in the selected neighborhood of a single voxel.
Section~\ref{sec-4} describes the
procedure of constructing our proposed test statistic, and an adaptive
method for selecting neighboring voxels. Section~\ref{sec-4-2} explores
the theoretical properties of our proposed test statistic. Section \ref
{sec-5} presents
simulation results. There, our methods are compared with FA-threshold
and Smooth-FA-threshold approaches. Section
\ref{sec-6} applies all approaches on real brain DTI data. Section
\ref
{sec-7} discusses our findings in this paper. Technical conditions and
proofs are given in a supplemental document.

\section{Background} \label{sec-2}

We begin with a brief introduction of DWI data, DTI data and existing
statistical models for estimating DTI data
from DWI data. Then, we summarize the associations among fiber tracts,
tissue types, water diffusivity and DT types. After that, we overview the
quantitative scalars, FA and RA, popularly used in fiber tracking algorithms.

\subsection{From DWI to DTI} \label{sec-2-1}

In this section we first give a brief introduction of the structures of
DWI and DTI data, where the former are acquired from the diffusion
weighted magnetic resonance experiment, while the latter are estimated
from the former based on some statistical model. Then, we summarize the
existing statistical models for estimating DTI data from DWI
data.

We assume that the DWI data over the brain of a given subject contain $N$
voxels, each of which
consists of diffusion-weighted measurements. Denote by $\phi_0$, $b$,
$\{(\phi_i, \bg_i)\}_{i=1}^r$, the acquired diffusion-weighted
measurements at a given voxel over the brain in a DWI experiment. Here,
the $i$th diffusion gradient $\bg_i = (g_{i,1}, g_{i,2}, g_{i,3})^T$,
with $\bg_i^T \bg_i = 1$, is chosen by the experimenter before the DWI
experiment starts, and serves as a scanning direction in the experiment
[\citet{Hasan2001}]; $b$ is the $b$-factor, whose value is determined by
a function of parameter settings in the DWI experiment [\citet
{StejskalTanner1965}, \citet{Moseley1990}, \citet{Anderson2001}]; both
$\bg_i$ and $b$ usually adopt the same values over all voxels in a DWI
experiment; $\phi_i$
denotes the diffusion attenuated signal, acquired on the $i$th
diffusion gradient $\bg_i$ at $b$; $\phi_0$ is the reference signal
obtained at $b= 0$; $\{\phi_i\}_{i=1}^r$ and $\phi_0$ compose the
responses of the DWI experiment for each voxel; $r$ is the number of
acquired attenuated signals for each voxel.\looseness=1

Accordingly, a single-voxel of the DTI data contains a $3 \times3$
symmetric, positive definite DT matrix,
\[
\bD= \lleft[\matrix{ D_{1,1} & D_{1,2} & D_{1,3}
\vspace*{2pt}
\cr
D_{1,2} & D_{2,2} & D_{2,3}
\vspace*{2pt}
\cr
D_{1,3} & D_{2,3} & D_{3,3} }
\rright],
\]
which
carries the intrinsic information of water diffusion in that particular voxel.
The elements of $\bD$ can be reorganized as a $6\times1$ vector
$\mathbf{d}= (D_{1,1}, D_{2,2}, D_{3,3}, D_{1,2},  D_{1,3}, D_{2,3})^T$.

The connections between DWI and DTI data are first investigated by the
seminal work of \citet{Basser1994}, in which the following
multivariate linear and nonlinear regression models are proposed.
For a single-voxel,
%
\begin{eqnarray}
&&\mbox{multivariate linear model:}\quad \log(\phi_i) = \log(
\phi_{0}) -b\bx_i^T\mathbf{d}+
\eps_i, \label {eq-21}
\\
&&\mbox{multivariate nonlinear model:} \quad\phi_i =
\phi_{0}\exp\bigl(-b\bx_i^T\mathbf{d}\bigr) +
\eta_i, \label{eq-22}
\end{eqnarray}
where $i = 1, \ldots, r$, $\bx_i = (g_{i,1}^2, g_{i,2}^2, g_{i,3}^2,
2g_{i,1}g_{i,2}, 2g_{i,1}g_{i,3}, 2g_{i,2}g_{i,3})^T$,
$\varepsilon_i$ and $\eta_i$ are random errors. $\mathbf{d}$ is then estimated
from either model by regression techniques.

A number of alternative statistical models as well as fitting
procedures, besides
\citet{Basser1994}, have been proposed to obtain sophisticated
DT estimates from DWI data concerning various aspects, such as
robustness, bias, non-Gaussian errors, spatial
smoothness, model validity, etc. Examples include \citet{Mangin2002},
\citet{Chang2005}, \citet{Salvador2005}, \citet{Heim2007}, \citet{Zhu2007},
\citet{Tabelow2008} and many others.

%
\begin{figure}

\includegraphics{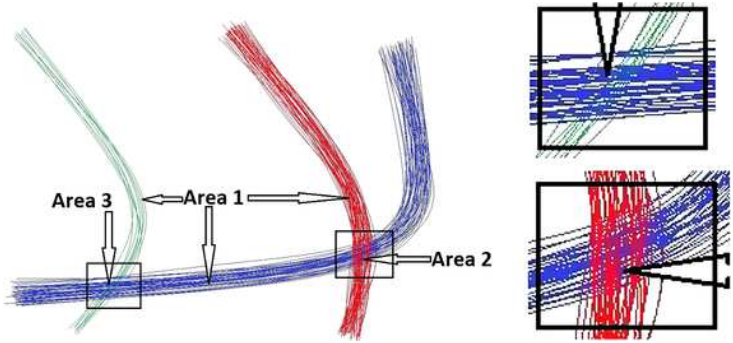}

\caption{An illustrative example of fiber tracts.}\label{Figure-Fiber-disp}
\end{figure}

\subsection{Fiber tracts, tissue types, water diffusivity and DT types}

The associations among fiber tracts, tissue types, water diffusivity
and DT types are summarized as follows. For voxels located in fiber
tracts, that is, white matter brain areas, water tends to present
higher diffusivity along the dominant orientation of fibers than that
in other orientations. DTs in these voxels are anisotropic,
characterized by the heterogeneity in lengths of axes in the
corresponding ellipsoid representations. In contrast, for voxels in
brain areas without fiber tracts, that is, grey matter areas, DTs are
isotropic. In these areas, the diffusivity of water in all orientations
is roughly the same. Therefore, the corresponding represented
ellipsoids are spherically shaped. The morphology of an anisotropic DT
can usually be classified into one of the three categories, namely,
prolate, oblate and nondegenerate, respectively, corresponding to brain
voxels located in uniquely orientated fiber tracts (\textit{Area} 1 in
Figure~\ref{Figure-Fiber-disp}), crossed fiber tracts with similar
intensities on two or more different orientations (\textit{Area} 2 in
Figure~\ref{Figure-Fiber-disp}, characterized by similar fiber
denseness for the red and blue bundles in their intersected parts) and
crossed fiber tracts with distinct intensities on different
orientations (\textit{Area} 3 in Figure~\ref{Figure-Fiber-disp},
characterized by the scenario that the denseness of blue tracts is
higher than that of green tracts in their intersected parts).

A vast number of tractography algorithms have been proposed in order to
reconstruct fiber tracts in the human brain based on DTI data. A list
of examples of these algorithms can be found in \citet{Conturo1999},
\citet{Gossl2002}, \citet{Xu2002}, \citet{Behrens2007}, \citet
{ODonnellWestin2007} and many others. We observe that most of these
approaches are founded on the derived voxel-wise scalar quantities,
such as FA and RA, which summarize/extract microstructural information
of water diffusion carried by DWI or DTI data.

\subsection{Quantitative scalars: FA and RA}\label{sec-2-2}

The fractional anisotropy (FA) and relative anisotropy (RA) [\citet
{BasserPierpaoli1996}] are quantitative scalars widely used by
neuroscientists to measure the water diffusivity in brain tissues and
construct algorithms for tracking fibers, since they are
computationally simple and invariant in
the choice of the laboratory coordinate system and
diffusion gradients. For a given voxel, FA and RA are
defined as
\[
\FA=\sqrt{\frac{3}{2}\frac{\sum_{k=1}^3\{\lambda_{(k)}-\ol
\lambda_{(\cdot)}\}^2}{\sum_{k=1}^3\lambda_{(k)}^2}},\qquad \RA =\frac{3}{\sqrt{2}}
\frac{\sqrt{\sum_{k=1}^3\{\lambda_{(k)}-\ol
\lambda_{(\cdot)}\}^2}}{\sum_{k=1}^3\lambda_{(k)}},
\]
where $\ol\lambda_{(\cdot)}
=\{\lambda_{(1)}+\lambda_{(2)}+\lambda_{(3)}\}/{3}$, with
$\lambda_{(3)}\ge\lambda_{(2)} \ge\lambda_{(1)}$ being the ordered
eigenvalues of the DT estimate.

Under the ideal but unrealistic assumption that the estimated DT is
noise free, $\RA=0$ and $\FA=0$ in isotropic voxels [i.e.,
$\lambda_{(1)} = \lambda_{(2)} =\lambda_{(3)}$], whereas $\RA=\sqrt{3}$
and $\FA=1$ in purely
anisotropic voxels [i.e.,
$\lambda_{(3)}\gg\lambda_{(2)} =\lambda_{(1)}$]. However, the noise
carried by the DWI data contaminates the DT estimates, and subsequently
introduces systematic
bias into the derived eigenvalues [\citet{PierpaoliBasser1996}].
Although these bias components have been investigated by numerical evaluations
[\citet{Jones2003}, \citet{LazarAlexander2003}] as well as in theory
[\citet{Zhu2007}], the magnitudes are sensitive to the distribution of
the noise in DWI experiments. Consequently, these bias components
introduce uncertainty into the constructed FA and RA. In practice,
brain areas with small but nonvanishing FA (RA) usually correspond to
grey matter areas, whereas those with large FA (RA) are typically areas
passed by fiber tracts. Some tractography algorithms are based on FA or
RA with thresholds (e.g., the tractography algorithm integrated in
\texttt{MedINRIA}, a publicly available software at
\url{http://www-sop.inria.fr/asclepios/software/MedINRIA}). The
thresholds, however, are usually manually chosen by investigators based
on their historical knowledge of the DTI data and the structure of the
human brain.

\begin{figure}

\includegraphics{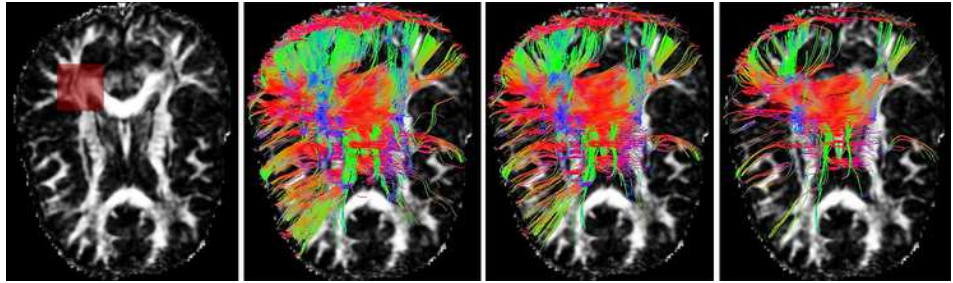}

\caption{Fiber tracking results by distinct FA thresholds. From left
to right:
FA map with a ROI highlighted; constructed fibers passing through
the ROI by FA thresholds $0.3$, $0.45$ and $0.6$,
respectively.}\label{Figure-1}\vspace*{-3pt}
\end{figure}

Figure~\ref{Figure-1} illustrates how
the fiber tracking results are affected by
distinct experiential thresholds of the FA, where the detailed
data information is given in Section~\ref{sec-6}.\vadjust{\goodbreak}
We use a region of interest (ROI) with $30\times30$ voxels, which is
highlighted by a red rectangle shown in the leftmost panel of Figure~\ref{Figure-1}.
The tracked fibers passing through the ROI using
distinct FA thresholding criteria, namely, $\FA> c$ with $c = 0.3,
0.45$ and $0.6$, are displayed in the remaining three panels of Figure
\ref{Figure-1}. All sub-figures in Figure~\ref{Figure-1} are
constructed by \verb"MedINRIA", where different colors stand for
different principal orientations of the corresponding DTs.
We observe that the fiber tracking results are sensitive to the FA
threshold, the choice of which, owing to the uncertainty of the bias
magnitudes in eigenvalue estimates, is not well supported in theory. In
other words, there is lack of a criterion to choose the threshold for
FA, and justify the goodness of the tracking results.

%
%

A more sophisticated approach based on FA is due to \citet{Zhu2006}, in
which the asymptotic distribution of a test statistic established on FA
is investigated. Therefore, it is capable of suggesting a threshold to
be used. This approach, however, is not applicable in our study, since
the theory needs the assumption that the number of gradients for every
brain voxel is diverging to infinity, that is, $r\to\infty$. In
contrast, in our study $r = 12$.

Another limitation of FA and RA with threshold approaches is that the
DTI data are typically spatially structured. FA and RA, however, are
defined as functions of the eigenvalues of the DT in every single
voxel. These approaches identify the existence of fibers in each voxel
while ignoring the spatial information, and therefore may diminish the
effectiveness in the downstream fiber tracking algorithms.

We observe that there exist approaches which accommodate the spatial
information in the stage of DT estimation, such as \citet{Heim2007} and
\citet{Tabelow2008}. Intuitively, by incorporating the spatial
information, these methods improve the DT estimation, and therefore the
accuracy of FA and RA. Hereafter, we refer to the FA based on DT
estimated by the approach in \citet{Tabelow2008} and \citet
{PolzehlTabelow2009} (implement in \texttt{R} package \texttt{dti}) as
\textit{Smooth-FA}. In our numerical studies, Smooth-FA with threshold,
namely, \textit{Smooth-FA-threshold}, is employed as an approach to
identify anisotropic brain voxels, and is compared with our proposed
methods.\looseness=-1\vadjust{\goodbreak}

\section{Statistical model on local eigenvalues} \label{sec-3}

We consider the DTI data of a given subject with $N$ voxels. In each
voxel $v \in\{1, \ldots, N\}$, denote by $\bD(v)$ (without confusion,
we concisely denote $\bD$) the DT estimated from a statistical model in
Section~\ref{sec-2-1}. Let $\lambda_{(3)}\ge\lambda_{(2)} \ge
\lambda_{(1)}$ be the ordered eigenvalues of $\bD$. In practice, these
eigenvalues are adopted to estimate the ordered true eigenvalues,
denoted by $\lambda_{(3)}^{*}\ge\lambda_{(2)}^{*} \ge\lambda_{(1)}^{*}$.

It has been demonstrated both in numerical studies and in theory that
$E\{\lambda_{(3)}\}>\lambda_{(3)}^{*}$ and $E\{\lambda_{(1)}\} <
\lambda_{(1)}^{*}$.\vspace*{1pt} Such a bias is actually caused by the sorting procedure in
the decomposition of $\bD$. In other words, we can postulate that one
of the eigenvalues of $\bD$, that is, $\lambda_k \in\{\lambda_{(k)}\dvtx  k
= 1,2,3\}$, is associated with $\lambda_{(k)}^{*}$ (but we don't know
which one it is). Here, ``associated'' means there exists a random error
$\varepsilon_k$ with $E(\varepsilon_k) = 0$ such that
%
\begin{equation}
\lambda_k = \lambda_{(k)}^{*} +
\varepsilon_k. \label{eq-33-added}
\end{equation}
However, $\lambda_{(k)}^{*}$ is estimated by $\lambda_{(k)}$ instead of
$\lambda_k$. Therefore, $E\{\lambda_{(3)}\}> E(\lambda_3) = \lambda_{(3)}^{*}$.
Similar arguments lead to $E\{\lambda_{(1)}\}< \lambda_{(1)}^{*}$. The magnitudes of the bias components are sensitive to the
distribution of the noise in DWI experiments, introducing uncertainty
into the quantitative scalars, such as FA and RA.

In our approach, instead of using the eigenvalues at a single voxel $v$
only, we choose a set of $n$ neighboring voxels located adjacent
to $v$. Denote by $\{\lambda_{j,(k)}(v)\dvtx  k=1,2,3\}_{j=1}^n$ [i.e.,
$\lambda_{j,(k)}$] the
eigenvalue estimates in the selected neighboring voxels, whose permuted
version associated with the set of true eigenvalues is denoted by $\{
\lambda_{j,k}(v)\dvtx  k=1,2,3\}_{j=1}^n$ (i.e.,
$\lambda_{j,k}$). Here, $j=1$
corresponds to voxel $v$, that is, $\lambda_{1,k} = \lambda_{k}$.

Similar in spirit to the random complete block design, we model
$E(\lambda_{j,k})$, the expected eigenvalue of the $j$th neighboring
voxel of $v$, as the addition of two components, namely, the true
eigenvalue $\lambda_{(k)}^{*}$ for voxel $v$ and the difference of
eigenvalues between voxel $v$ and its $j$th neighboring voxel, denoted
by $\beta_j$. That is, $E(\lambda_{j,k}) = \lambda_{(k)}^{*} + \beta_j$, which leads to the additive model,
%
\begin{equation}
\lambda_{j,k}= \lambda_{(k)}^{*}+
\beta_{j}+\eps_{j,k}, \label{eq-33}
\end{equation}
where the eigenvalues $\{\lambda_{(k)}^{*}\dvtx  k=1,2,3\}$ in voxel $v$ are
of primary interest and therefore treated as the treatment effects,
whereas the differences of eigenvalues between voxel $v$ and its
neighboring voxels $\{\beta_j\}_{j=1}^n$ are the blocking effects and
serve as nuisance parameters; $\{\varepsilon_{j,k}\dvtx  k=1,2,3\}_{j=1}^n$ are
random errors. Clearly, $\beta_1 = 0$, such that (\ref{eq-33}) complies
with (\ref{eq-33-added}) when $j = 1$.

Nonetheless, as discussed above, based on the DT estimates, we can only
obtain $\{\lambda_{j,(k)}, k = 1,2,3\}$, the ordered version of $\{
\lambda_{j,k}, k = 1,2,3\}$. In other words, the values of $\lambda_{j,k}$ are not available in practice. Therefore, the well-developed
techniques for the additive model in a complete
block design are not directly applicable in analyzing model (\ref
{eq-33}). We borrow its basic idea and integrate the corresponding
contrast test statistic as one of the main parts in our proposed test
statistic. The technical properties of our proposed test statistic in
isotropically diffused DT voxels, that is, $
\lambda_{(3)}^{*} = \lambda_{(2)}^{*} = \lambda_{(1)}^{*}$, then rely
on the fact that
$\lambda_{j,(k)}= \lambda_{(1)}^{*}+\beta_{j}+\eps_{j,(k)}$.

\section{Local test based on neighboring eigenvalues} \label{sec-4}

\subsection{Test of hypotheses} \label{sec-4-1}

Classifying the tissue types for a particular brain area
plays a key role in the downstream fiber tracking algorithms. In voxels
without fibers, water tends to diffuse in an isotropic manner,
characterized by the eigenvalue property
$\lambda_{(3)}^{*}\approx
\lambda_{(2)}^{*}\approx\lambda_{(1)}^{*}$. In contrast, water
in voxels passed by fiber tracts usually presents anisotropic
diffusion. DTs in these voxels typically have three possible
morphologies, namely, prolate $\lambda_{(3)}^{*} >
\lambda_{(2)}^{*} \approx\lambda_{(1)}^{*}$,\vspace*{1pt} oblate
$\lambda_{(3)}^{*} \approx\lambda_{(2)}^{*} >
\lambda_{(1)}^{*}$ and nondegenerate\vspace*{1pt} $\lambda_{(3)}^{*} >
\lambda_{(2)}^{*} > \lambda_{(1)}^{*}$, respectively, corresponding to
brain voxels located in uniquely oriented fiber areas, crossed fiber
areas with roughly the same fiber intensities and crossed fiber areas
with distinct fiber intensities. In summary, anisotropy
is characterized by either $\lambda_{(3)}^{*} > \lambda_{(2)}^{*}$ or
$\lambda_{(2)}^{*} > \lambda_{(1)}^{*}$. Therefore, we consider
the hypothesis testing problem,
%
\begin{eqnarray}
H_0\dvtx \lambda_{(3)}^{*}=\lambda_{(2)}^{*}
= \lambda_{(1)}^{*} \quad\mbox{vs.}\quad H_1\dvtx
\lambda_{(3)}^{*}>\lambda_{(2)}^{*} \quad
\mathrm{or}\quad \lambda_{(2)}^{*}>\lambda_{(1)}^{*}
\label{eq-41}
\end{eqnarray}
for each voxel $v$. This testing problem can be reformulated as a
form of contrast test,
%
\begin{eqnarray}\qquad
H_0\dvtx A\blambda^{*} = \bzero \quad\mbox{vs.}\quad
H_1\dvtx A(l,\cdot)\blambda^{*} > 0\qquad \mbox{for $l=1$, or,
\ldots, or $\rankA $}, \label{eq-42}
\end{eqnarray}
where $\blambda^{*}=(\lambda_{(3)}^{*},
\lambda_{(2)}^{*}, \lambda_{(1)}^{*})^T$, $A$ is a full row rank
matrix with $\rank(A) = \rankA$ and $\sum_{l_2=1}^3 A(l_1,l_2) =
0$ for $l_1 = 1,\ldots,\rankA$. There exists more than one possible
choice of
$A$, such that hypothesis testing problems (\ref{eq-41}) and
(\ref{eq-42}) are equivalent. For example,
%
\begin{equation}
A=\left[\matrix{ 1 & -1 & 0
\cr
0 & 1 & -1}\right], \label{eq-43}
\end{equation}
which is adopted in our numerical studies. Clearly, $\rankA=2$.

To test (\ref{eq-42}), for each voxel $v$, we
propose the test statistic $\chiK$, represented by
%
\begin{equation}
\chiK=n (\bU_n-\hat\btheta_{\bU_n})^T \hat
\Sigma^{-1} (\bU_n-\hat\btheta_{\bU_n}), \label{eq-44}
\end{equation}
where $\bU_n = ({A\ol{\blambda}_{\cdot}}/{\sqrt{\MSE}})\cdot
\sqrt{\ol
{S_{\cdot}^2}}$;
$\ol{\blambda}_{\cdot} = (\ol\lambda_{\cdot,(3)}, \ol
\lambda_{\cdot,(2)}, \ol\lambda_{\cdot,(1)})^T$; $\ol
\lambda_{\cdot,(k)}=\break\sum_{j=1}^n\lambda_{j,(k)}/  n$,
$k=1,2,3$; $\ol{S_{\cdot}^2}$ is the mean of $S_{j}^2$ over the
neighboring voxels of $v$;
$S_{j}^2=\sum_{k=1}^3\{\lambda_{j,(k)}-\ol\lambda_{j,
(\cdot)}\}^2/2$, $j=1,\ldots, n$; $\MSE=\sum_{j=1}^{n}
\sum_{k=1}^3 \{\lambda_{j,(k)}-\ol
\lambda_{\cdot,(k)}- \ol\lambda_{j,(\cdot)}+\ol\lambda_{\cdot
,(\cdot)}\}^2/\{2(n-1)\}$; let $\hat\calV_0$\vadjust{\goodbreak} be an estimate of
$\calV_0$, the set of isotropic voxels over the entire brain;
$\hat\btheta_{\bU_n}$ denotes the sample median of $\bU_n$ over
$\hat\calV_0$; $\hat\Sigma$ denotes the sample covariance of
$\sqrt{n}\bU_n$ over $\hat\calV_0$. $\hat\calV_0$ is
obtained from the following iteration
steps:
\begin{longlist}[(iii)]
\item[(i)] Evaluate $\bU_n$'s over all voxels of interest. For some
pre-given significant level $\alpha$, let $\chi_{\rankA;1-\alpha}^2$
be the $(1-\alpha)$th quantile of the chi-square distribution with
$\rankA$
degrees of freedom.
\item[(ii)] Let $\hat\calV_{0,0}$ include all voxels of interest.
\item[(iii)] In the $s$th iteration ($s=1, 2, \ldots$), let $\tilde
\chiK$ denote the statistic computed from~(\ref{eq-44}) based on
$\hat\calV_{0,s-1}$, and let $\chiK= c\tilde\chiK$.
\item[(iv)] Include voxel $v$ in $\hat\calV_{0,s}$ if the corresponding
statistic $\chiK<\chi_{\rankA;1-\alpha}^2$.
\item[(v)] Repeat steps (iii) and (iv) until $\hat\btheta_{\bU_n}$ converges.
\end{longlist}

In step (iii) above, the constant $c$ is to correct the bias in the
evaluation of $\chiK$. This is because in the $s$th iteration, $\hat
\btheta_{\bU_n}$ in $\chiK$ is constructed by only using voxels in
$\hat\calV_{0,s-1}$, which includes voxels with $\chiK$ capped by
$\chi_{\rankA, 1-\alpha}^2$ in the $(s-1)$th iteration. The value of $c$ is
derived as follows:
\begin{eqnarray*}
E\bigl\{\tilde\chiK\ID\bigl(c\tilde\chiK< \chi_{\rankA;1-\alpha}^2
\bigr)\bigr\} &\approx& \frac{1}{|\hat\calV_{0,s-1}|}\sum_{v: c\tilde\chiK<
\chi
_{\rankA;1-\alpha}^2 }n(
\bU_n-\hat\btheta_{\bU_n})^T\hat
\Sigma^{-1}(\bU_n-\hat \btheta_{\bU_n})
\cr
&\approx&
\frac{1}{|\hat\calV_{0,s-1}|} \sum_{v\in\hat\calV_{0,s-1}} \bigl\{\sqrt{n}(
\bU_n-\ol \bU_{\cdot})\bigr\}^T\hat
\Sigma^{-1}\bigl\{\sqrt{n}(\bU_n-\ol\bU_{\cdot
})\bigr\}
\\
&=& \frac{|\hat\calV_{0,s-1}|-1}{|\hat\calV_{0,s-1}|}\rankA \approx \rankA,
\end{eqnarray*}
where ``$=$'' is followed by the fact that in step (iii), $\hat\Sigma$
is the sample covariance of $\sqrt{n}\bU_n$ over
$\hat\calV_{0,s-1}$; $\ID(\cdot)$ is the indicator function. Therefore,
%
\begin{equation}
E\bigl\{\chiK\ID\bigl(\chiK< \chi_{\rankA;1-\alpha}^2\bigr)\bigr\}=E
\bigl\{c\tilde\chiK \ID\bigl(c\tilde\chiK< \chi_{\rankA;1-\alpha}^2\bigr)
\bigr\} \approx c\rankA. \label{eq-45}
\end{equation}
On the other hand, according to Theorem~\ref{Thm-1} in Section \ref
{sec-4-2}, $\chiK$ is
approximately $\chi_{\rankA}^2$ distributed,
%
\begin{equation}
E\bigl\{\chiK\ID\bigl(\chiK< \chi_{\rankA;1-\alpha}^2\bigr)\bigr\}
\approx \int_0^{\chi_{\rankA;1-\alpha}^2} \frac{t^{\rankA/2}}{2^{\rankA/2}\Gamma(\rankA/2)}e^{-t/2}
\,dt. \label{eq-46}
\end{equation}
Combining (\ref{eq-45}) and (\ref{eq-46}), we set $c =
\frac{1}{\rankA}\int_0^{\chi_{\rankA;1-\alpha}^2}
\frac{t^{\rankA/2}}{2^{\rankA/2}\Gamma(\rankA/2)}e^{-t/2}\,dt$.

\begin{remark} \label{remark-1}
We now make some remarks concerning the construction of the
statistic in (\ref{eq-44}). The main term $\bU_n$ in $\chiK$
consists of
two parts:
\begin{itemize}
\item The first part ${A\ol{\blambda}_{\cdot}}/{\sqrt{\MSE}}$
mimics the
$t$ statistic of the contrast test
for the additive model in a complete block design. Referring to the
proofs of Theorems~\ref{Thm-1} and~\ref{Thm-2},\vadjust{\goodbreak} under certain
regularity conditions, it approaches infinity with rate $\sqrt{n}$,
when there are significant differences among $\{\lambda_{(k)}^{*}\dvtx k=1,2,3\}$,
whereas it converges to a fixed constant when
$\lambda_{(3)}^{*} = \lambda_{(2)}^{*} = \lambda_{(1)}^{*}$.
Therefore, it
has good statistical power
in identifying the differences among $\{\lambda_{(k)}^{*}\dvtx k=1,2,3\}$,
when model~(\ref{eq-33}) is valid.
\item The second part
$\sqrt{\ol{S_{\cdot}^2}}$ is added to increase the power of the test
statistic $\chiK$ on the boundary of fibers. For any
voxel on the boundary of fibers, in the sense that its selected neighborhood
contains voxels belonging to both fiber and nonfiber areas, the
assumption that the collected eigenvalues in the selected neighboring
voxels follow model (\ref{eq-33}) may not be appropriate. In this case,
$\sqrt{\MSE}$ tends to inflate and, consequently, the statistical power
of the
first part ${A\ol{\blambda}_{\cdot}}/{\sqrt{\MSE}}$ is limited. The
second part then counteracts the effect of $\sqrt{\MSE}$, and
therefore is
particularly useful when the resolution of the DTI data is limited.
\item Furthermore, $\widehat{\btheta}_{\bU_n}$, the sample median
instead of sample mean of $\bU_n$, is adopted just to ensure the
robustness of the approach.
\end{itemize}
\end{remark}

\begin{remark} \label{remark-2}
In this paper we establish testing procedures for identifying the brain
areas with fiber tracts. However, for voxels located in fiber tracts,
their DTs have three possible morphologies, namely, prolate, oblate,
and nondegenerate, respectively, corresponding to eigenvalue properties
$\lambda_{(3)}^*>\lambda_{(2)}^*\approx\lambda_{(1)}^*$,
$\lambda_{(3)}^*\approx\lambda_{(2)}^*>\lambda_{(1)}^*$, and $\lambda_{(3)}^*>\lambda_{(2)}^*>\lambda_{(1)}^*$. One may also implement
similar testing procedures as those in this paper to further classify
these three possibilities. We leave the details out for presentational brevity.

\end{remark}

\subsection{Adaptive selection of neighborhood} \label{sec-4-3}

Following Sections~\ref{sec-4-1} and~\ref{sec-4-2},
the asymptotic theories for $\chiK$ need that the
neighborhood size $n$ is large and that model~(\ref{eq-33}) is
satisfied. From the experimental point of view, as long as the
resolution in a DWI experiment is sufficiently good, such that the
proportion of voxels
located on the boundary of fibers over the entire brain shrinks, we can
simply employ a fix-shaped
neighborhood in the construction of $\chiK$ (e.g., choose the
neighborhood as a fixed cube). However, for experiments with limited
resolution, a fix-shaped neighborhood may not be a good choice, because
in this case, the assumption that the eigenvalues in the neighboring
voxels follow model (\ref{eq-33}) may not be well satisfied, in order
to ensure $n$ large required by Theorem~\ref{Thm-1}. Such a problem is
particularly severe for voxels located on the boundary of fibers.
Therefore, development of a varied neighborhood is necessary.

We propose an adaptive method to select the neighboring voxels based on
the philosophy below.
First, the adjacent voxels of $v$ should have a better chance to be
selected as neighboring voxels than those far away.
Second, to ensure the validity of model
(\ref{eq-33}), if the tensor in $v$ is isotropic, the selected
neighborhood should mainly\vadjust{\goodbreak} consist of voxels with isotropic tensors.
Likewise, if the tensor in
$v$ is anisotropic, the selected neighborhood should be in
favor of voxels with anisotropic tensors. Therefore, we incorporate the
physical distances and
similarity measures of DTs between voxel $v$ and its nearby
neighbors to establish the criteria for selecting neighboring voxels.

For each voxel $v$ and fixed number $n$ of neighboring voxels, we
summarize our
proposed adaptive neighborhood selection approach as follows:
\begin{longlist}[(1)]
\item[(1)] Fix a cube-shaped domain centered at $v$ with reasonably
large size $x\times y \times z$, whose voxels are
candidates. Here $x,y,z$ are integers and $xyz\ge n$.

\item[(2)] Define the similarity score function $f$ between voxel $v$ and
its neighboring candidate $v_l$, $l = 1,\ldots,xyz$, as
\[
f(v, v_l) = d_{\bD}\bigl(\bD(v),\bD(v_l)
\bigr)\exp\bigl\{C\cdot d_p(v,v_l)\bigr\},
\]
where $d_p(v,v_l)$ denotes the physical distance between voxels $v$
and $v_l$; $d_{\bD}(\bD(v),\break  \bD(v_l)) =
\sqrt{\trace[\{\bD(v)-\bD(v_l)\}^2]}$ is a measurement of the
diffusion similarity between tensors in voxels $v$ and $v_l$
[\citet{Alexander1999}]; $C\ge0$ is added to balance the
contribution of $d_p(v, v_l)$ and $d_{\bD}(\bD(v), \bD(v_l))$.

\item[(3)] Select $n$ voxels with the lowest $f$ values as the
neighboring voxels.
\end{longlist}

We observe that $C$ in $f$ is adopted to balance the contribution of
$d_{\bD}(\bD(v), \break \bD(v_l))$ and $d_p(v,v_l)$. When the resolution of
the DTI data is high, $d_p(v,v_l)$ is close to 0. Consequently, for any
fixed $C$, $\exp\{C\cdot d_p(v,v_l)\}$ approaches the constant 1.
Therefore, for high resolution DTI data, the proposed approach is not
sensitive to the choice of $C$. Throughout our numerical studies, we
fix $C = 0.1$.

We would like to point out that the proposed approach above is similar
in spirit to the adaptive approaches in \citet{Tabelow2008} and \citet
{Li2011}, where iterative testing procedures are used to adaptively
control the contribution of neighboring voxels in their proposed
algorithms. Compared with their approaches, our approach is
computationally more economic. The effectiveness of our proposed
approach above has been demonstrated in \citet{Yu2009} by simulation studies.

We summarize our proposed procedure of constructing $\chiK$ in the
supplemental document [\citet{Yu2012}].

\section{Theoretical properties} \label{sec-4-2}

In this section we explore the theoretical properties of our
proposed test statistic $\chiK$. The technical details are given in the
supplemental document [\citet{Yu2012}].
Theorem~\ref{Thm-1} below establishes the asymptotic null distribution of
$\chiK$, when the number $n$ of neighboring voxels is
large.

\begin{theorem} \label{Thm-1}
Assume model (\ref{eq-33}) and Condition \textup{A} in the
supplemental document. Then
for $\chiK$ defined in (\ref{eq-44}),
under the null hypothesis in (\ref{eq-42}), as $n\to\infty$,
\[
\chiK\conD\chi_{\rankA}^2.\vadjust{\goodbreak}
\]
\end{theorem}

We would like to point out that the construction of $\chiK$ and the
theoretical derivations of Theorem~\ref{Thm-1} are nontrivial and
challenging. Following the discussion in Section
\ref{sec-3}, for each voxel $v$, we postulate that there exist
unobservable one-to-one
correspondences, that is, $\{(\lambda_{k}, \lambda_{(k)}^{*})\dvtx k =
1,2,3\}$, between the estimated and true eigenvalues. The ordered
eigenvalue estimates $\lambda_{(3)}\ge\lambda_{(2)}\ge\lambda_{(1)}$,
however, are neither
unbiased estimates for
$\lambda_{(3)}^{*}\ge\lambda_{(2)}^{*}\ge\lambda_{(1)}^{*}$,
nor independent. We address these bias components in the construction
of $\chiK$ and the corresponding proof of Theorem~\ref{Thm-1} based on
the intuition as follows. Referring to model (\ref{eq-33}), for an
isotropic voxel $v$, the collected neighboring voxels can be modeled as
$\lambda_{j,(k)}= \lambda_{(1)}^{*}+\beta_{j}+\eps_{j,(k)}$. As such,
the bias components of eigenvalue estimates in the neighboring voxels
of $v$ are carried by $\eps_{j,(k)}$, whose effects in our test
statistic are counteracted by $\hat\btheta_{\bU_n}$ constructed based
on spatial information of the entire brain.

To appreciate the discriminating power of $\chiK$ in
the identification of aniso\-tropic brain areas, the asymptotic
power of $\chiK$ is established in Theorem~\ref{Thm-2} below.

\begin{theorem} \label{Thm-2}
Assume model (\ref{eq-33}) and Condition \textup{A} in the
supplemental document. Then for voxel $v$, under the fixed
alternative $H_1$ in (\ref{eq-42}), as $n\to\infty$,
\[
n^{-1}\chiK\conP M,
\]
where $M$ is given by \textup{(A.4.3)} in the supplemental document.
\end{theorem}

Theorem~\ref{Thm-2} shows that as long as $\bg(\tilde\ba(v)) \ne
\bg
(\bb)$, $M>0$ and $\chiK\conP+\infty$
at rate~$n$, under the fixed alternative
$H_1$. Here, $\bg(\cdot)$ is
defined by (A.3.2) in the supplemental document; $\tilde\ba(v) = (E\{
2S_1^2(v)\},$
$E\{\lambda_{1,(3)}(v)\}, E\{\lambda_{1,(2)}(v)\},\break
E\{\lambda_{1,(1)}(v)\})^T$; $\bb= (E(2S_{\varepsilon}^2),
E\{\varepsilon_{1,(3)}(1)\}, E\{\varepsilon_{1,(2)}(1)\},E\{\varepsilon_{1,(1)}(1)\})^T$; $S_1^2(v)$
and $S_{\varepsilon}^2$ are, respectively, the
sample variances of $\{\lambda_{1,(k)}(v), k = 1,2,3\}$ and $\{
\varepsilon_{1,(k)}(1), k = 1,2,3\}$. Thus, under the fixed alternative, the power
of our proposed test statistic $\chiK$ tends to 1 except in rare
situations. Corollary~\ref{corr-1} below gives one specific example of
$M>0$. The proof is straightforward and omitted.

%
%
\begin{figure}

\includegraphics{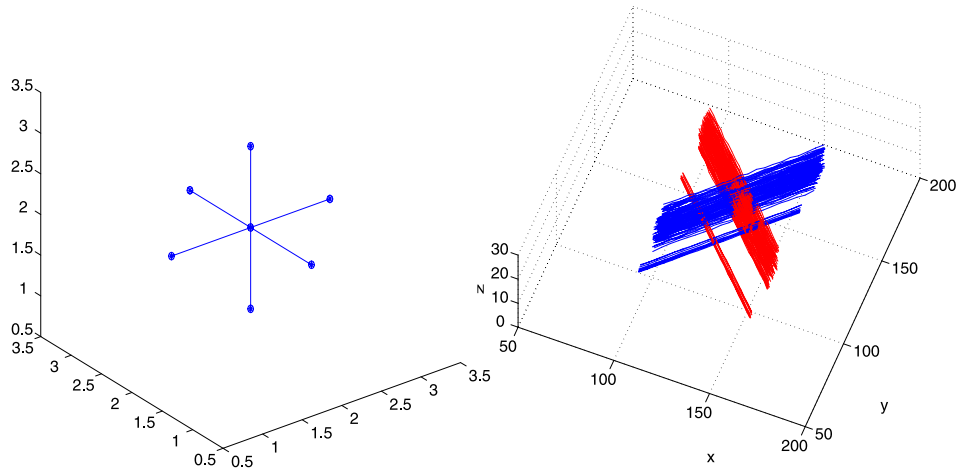}

\caption{Left panel---neighbors of a voxel used in the $\FDR_L$
procedure. Right panel---gometry of the simulated brain.}
\label{Figure-2}
\end{figure}

\begin{corollary} \label{corr-1}
Assume conditions in Theorem~\ref{Thm-2}. Suppose that
$\varepsilon_{j,k}$ has a symmetric distribution about $0$, that is,
$\varepsilon_{j,k}$ has the same distribution as $-\varepsilon_{j,k}$, and
$E\{\lambda_{1,(3)}\}- E\{\lambda_{1,(2)}\} \ne
E\{\lambda_{1,(2)}\}- E\{\lambda_{1,(1)}\}$. Then $M>0$.
\end{corollary}

\section{Simulation study} \label{sec-5}

\subsection{Basic settings for numerical work} \label{sec-5-1}

Since in a real DTI data set, the number of voxels, each of which
corresponds to a hypothesis test, is typically large, false discovery
rate (FDR) techniques\vadjust{\goodbreak} [\citet{BenjaminiHochberg1995},
\citet{Storey2002},
\citet{Storey2004}, \citet{Zhang2011}] are incorporated in our numerical
works to control the error rates. Two FDR procedures are employed in
our study, namely, the
conventional FDR procedure by \citet{Storey2002} and the $\FDR_L$
procedure by
\citet{Zhang2011}, which is capable of capturing the spatial information
in imaging data. A short summary of the $\FDR_L$ procedure is provided
in the supplemental document [\citet{Yu2012}].

Some settings of parameters throughout our numerical works are given as
follows. For $\FDR$ and $\FDR_L$
procedures, set false discovery control level as 0.01 and tuning
parameter $\lambda= 0.2$. The neighborhood for the $\FDR_L$ procedure
is set as
the nearest 7 voxels shown in the left panel of Figure
\ref{Figure-2}. When evaluating $\chiK$ for each voxel, $A$ is given by
(\ref{eq-43}). Apply the adaptive
neighborhood selection approach proposed in Section~\ref{sec-4-3} to
select neighboring
voxels, where the domain for the candidate neighboring
voxels is set as a $5\times5\times3$ cube centered at $v$. There,
$n=25$ voxels
are selected based upon $f$ with $C = 0.1$. The choice of $n$ is
referred to in the results in Section~\ref{sec-5-2-2}: when $n=25$, the
sampling distribution of $\chiK$ agrees reasonably well with the $\chi^2$ distribution.

\subsection{Data simulation} \label{sec-5-2-1}

We simulate several sets of DWI data over
the entire brain. For each set, a 3D imaging space with the same brain
areas as the real DWI data in Section~\ref{sec-6} is simulated, where
fiber tracts are simulated to have the geometrical structure displayed
in the right panel of Figure~\ref{Figure-2}. DTs are simulated
according to the locations of voxels in the imaging space. To this end,
four distinct sets of eigenvalues, namely, $[0.7, 0.7, 0.7]$, $[1.0,
0.55, 0.55]$, $[0.8, 0.8, 0.5]$ and $[0.9, 0.7, 0.5]$ (units: $10^{-3}
\mbox{mm}^2/\mathrm{s}$), are\break adopted to, respectively, simulate DTs with
morphologies of \textit{isotropic} (nonfiber areas), \textit{prolate}
(single blue and red bundles), \textit{oblate} (intersected areas of
blue bundles) and \textit{nondegenerate} (intersected areas between
blue and red bundles), such that all simulated DTs share the same mean
diffusivity $\bar\lambda_{(\cdot)}^{*} = 0.7 \times10^{-3} \mbox{mm}^2/\mathrm{s}$,
a typical value in real human brains [\citet{Pierpaoli1996},
\citet{Anderson2001}].

Since the acquired attenuated signal intensity, $\phi_i(v)$, at each
voxel $v$ and gradient $\bg_i$ in real DWI data is typically generated
by the square-root of the sum of squares of two random numbers in the
DWI experiment [\citet{Henkelman1985}, \citet{Salvador2005}, \citet
{Zhu2007}], we simulate a reference signal ($i=0$) and $r=12$ diffusion
attenuated signals ($i=1,\ldots,12$) in each $v$ [$=(v_x,v_y,v_z)$] as
\[
\phi_i(v) = \sqrt{\bigl[\phi_0^*(v)\exp\bigl
\{-b\bg_i^T\bD^*(v)\bg_i\bigr\} +
\varepsilon_{i,x}(v)\bigr]^2+\varepsilon_{i,y}^2(v)},
\]
where $\phi_{0}^*(v) = 1200$ when $v_x \in(0, 128]$, $\phi_0^*(v)=1800$ when $v_x\in(128, 256]$; the $b$ factor $b = 1000$ when
$i > 0$, $b=0$ when $i = 0$; diffusion gradients $\{\bg_i\dvtx  i=1,\ldots,
12\}$ are adopted from the real DWI data in Section~\ref{sec-6}; $\bD^*(v) = Q(v)\Lambda^{*}(v)Q^T(v)$;
$\Lambda^{*}(v) = \diag(\lambda_{(3)}^{*}(v),
\lambda_{(2)}^{*}(v), \lambda_{(1)}^{*}(v))$; $Q(v)$ is a $3\times
3$ orthogonal matrix whose column vectors are composed of the
eigenvectors of the simulated $\bD^*(v)$,
\[
Q(v)=\left[\matrix{ 1/\sqrt{2}& 1/\sqrt{2} &0
\cr
-1/\sqrt{2} & 1/\sqrt{2} &0
\cr
0&0&1 }\right]\quad \mbox{or}\quad Q(v)=\left[\matrix{ 1/\sqrt{2}& -1/\sqrt{2} &0
\cr
1/\sqrt{2} & 1/
\sqrt{2} &0
\cr
0&0&1 }\right].
\]
Clearly, the former corresponds to the red and the corresponding
parallel narrow blue bundles in the right panel of Figure
\ref{Figure-2}, while the latter models the other two blue bundles.
The random errors $\eps_{i,x}(v)$ and $\eps_{i,y}(v)$ are simulated as
independent and normally distributed with variance $\sigma^2$, which is
varied to provide signal to noise ratios (SNRs), where SNR = $\phi_0^*(v)/\sigma$.
We examine four distinct SNRs, \{5, 10, 15, 20\}, each
corresponding to one set of the simulated DWI data.

\subsection{\texorpdfstring{Agreement between $\chi^2$ distribution and $\chiK$}
{Agreement between chi2 distribution and K}} \label{sec-5-2-2}

With the DWI data sets simulated in Section~\ref{sec-5-2-1}, we
estimate the DT in each voxel by regression model (\ref{eq-21}) and
the corresponding eigenvalues by Schur decomposition. For each DWI data
set, two sets (I and II) of
$\chiK$ are constructed according to different settings of the adaptive
neighborhood selection approach in Section~\ref{sec-4-3}. In
particular, for
\textit{Set} I, the size of the candidate neighborhood
and the number of selected neighboring voxels are, respectively, chosen as
$5\times5\times3$ and $n=25$, while those for
\textit{Set} II as $11\times11 \times3$ and $n=81$.
Other settings are given in Section~\ref{sec-5-1}.

\begin{figure}

\includegraphics{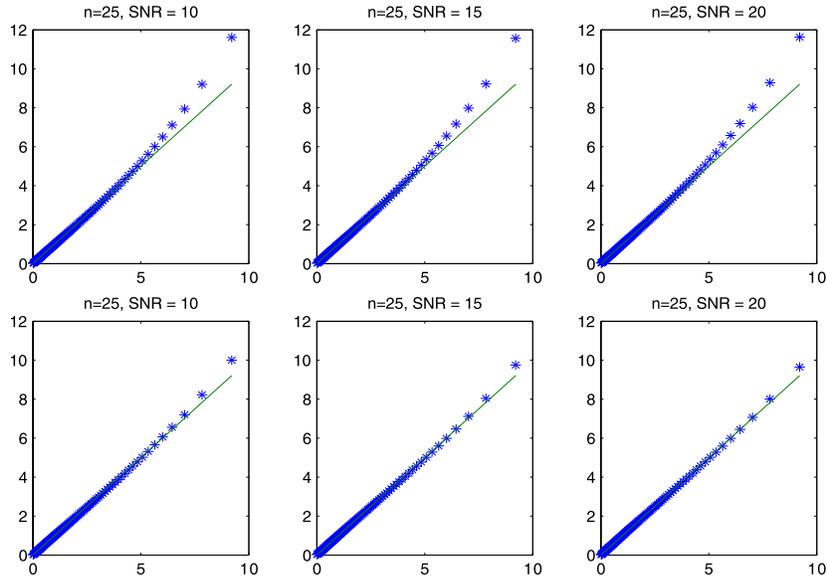}

\caption{Empirical percentiles of $\chiK$ ($y$-axis) versus
percentiles of
$\chi_{\rankA}^2$ distribution ($x$-axis). Top panels---candidate cubic
$5\times5\times3$, $n=25$. Bottom panels---candidate cubic $11\times
11\times3$, $n=81$. From left to right panels: $\SNR$ = $10, 15$ and
$20$. Solid line---the $45$
degree reference line.} \label{Figure-3}\vspace*{-3pt}
\end{figure}

For each simulated data set, we collect all $\chiK$'s whose
corresponding voxels are located inside the simulated nonfiber areas.
The QQ plots of the (1st to 99th) percentiles of these $\chiK$'s
against those of the $\chi_{\rankA}^2$ distribution are
displayed in Figure~\ref{Figure-3}, with top panels based on
\textit{Set} I, bottom panels on \textit{Set} II. The left, middle and right
panels correspond to $\SNR= 10, 15$ and $20$, respectively. Results
in Figure~\ref{Figure-3} demonstrate that the sampling
distributions of $\chiK$, under both $n=25$ and $n = 81$,
agree reasonably well with the $\chi^2$ distribution.

%
%

\subsection{Receiver operating characteristic curve} \label{sec-5-2-3-added}

We compare our methods ($\chiK$-FDR and $\chiK$-$\FDR_L$) with FA with
threshold (i.e., \textit{FA-threshold}) and Smooth-FA with threshold
(i.e. \textit{Smooth-FA-threshold}) approaches by the receiver
operating characteristic (ROC) curve, a widely adopted statistical tool
for evaluating the accuracy of continuous diagnostic tests [\citet{Pepe2003}].

Let $\{T_1, \ldots, T_N\}$ be the set of statistics for all voxels over
the entire brain, where $T_v$ for any voxel $v\in\{1,\ldots, N\}$ is
the FA or Smooth-FA value, if the FA-threshold or Smooth-FA-threshold
approach is used; is the $p$-value based on $\chiK$, if the $\chiK$-FDR
approach is used; is the $\tilde p$-value if the $\chiK$-$\FDR_L$
approach is used, where $\tilde p$ stands for the median smoothed
p-value of $\chiK$ [\citet{Zhang2011}]. For any given threshold $t$, if
we classify a voxel $v$ as anisotropic based on $T_v\in R(t)$,
where
$R(t)=\{T_v\dvtx  T_v\ge t, v = 1,\ldots,N\}$ when $T_v$ is the FA or
Smooth-FA value; $R(t) = \{T_v\dvtx  T_v\le t, v = 1,\ldots, N\}$ when $T_v$
is the $p$- or $\tilde p$-value, then,
\begin{eqnarray*}
&&\mbox{sensitivity:}\quad  \operatorname{se}(t) \equiv\frac{\sum_{v=1}^N I\{T_v \in R(t),
H_1 \mbox{ is true} \}}{|\calV_0|},
\\
&&\mbox{specificity:}\quad  \operatorname{sp}(t) \equiv\frac{\sum_{v=1}^N I\{T_v \notin
R(t), H_0 \mbox{ is true}\}}{N-|\calV_0|},
\end{eqnarray*}
where $|\calV_0|$ is the number of isotropic voxels over the entire
brain. The ROC curve is then constructed as the 2D curve ($\operatorname{se}(t)$, $1-\operatorname{sp}(t)$)
when $t$ ranges between 0 and 1. The area under the ROC curve (AUC) is
a popularly adopted measure of the accuracy of the test. More
precisely, AUC is ranging between 0 and 1, and the larger the AUC, the
better the method.

%
\begin{figure}

\includegraphics{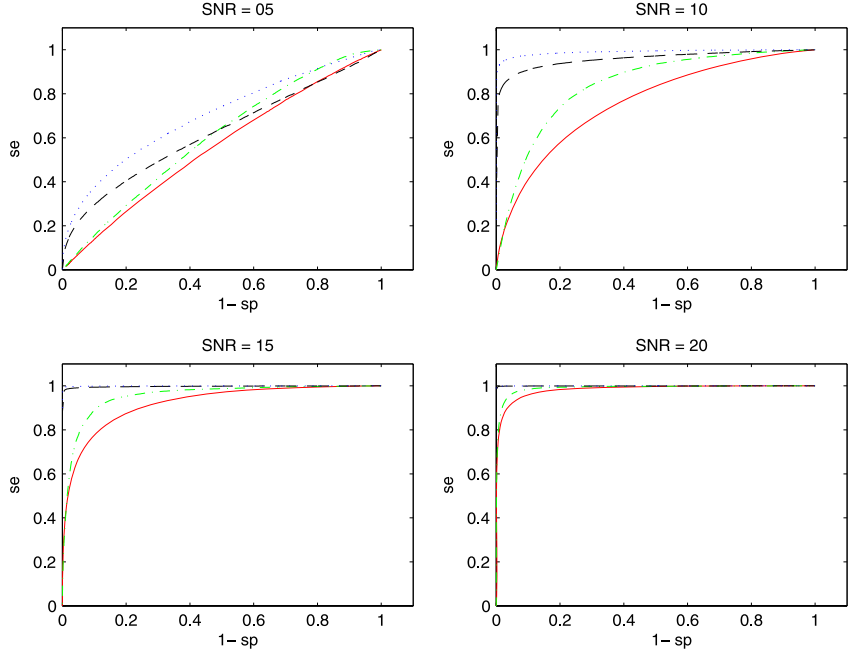}

\caption{ROC curves for SNR = 5, 10, 15 and 20. Dashed black
line---$p$-value of $\chiK$;
dotted blue line---$\tilde p$-value of $\chiK$; solid red line---FA;
dash-dotted green line---Smooth-FA.} \label{Figure-ROC}
\end{figure}

Following Sections~\ref{sec-5-1} and~\ref{sec-5-2-1}, the ROC curves
for FA, Smooth-FA, $p$ and $\tilde p$ are displayed in Figure \ref
{Figure-ROC} for data sets with $\operatorname{SNR} = 5, 10, 15$ and 20, respectively. It
has been seen from Figure~\ref{Figure-ROC} that the ROC curves of
$\tilde p$ (dotted blue lines) and~$p$ (dashed black lines) are
consistently located above those of FA (solid red lines) and Smooth-FA
(dash-dotted green lines) for all simulated data sets, indicating that
$\chiK$-FDR and $\chiK$-$\FDR_L$ are capable of achieving better
classification accuracy than FA-threshold and Smooth-FA-threshold
approaches. Furthermore, the AUCs of $\tilde p$ in all examined data
sets are superior to those of $p$, suggesting that $\chiK$-$\FDR_L$
performs the best among all four approaches.

\subsection{Test results} \label{sec-5-2-3}

In this section we present our test results in simulated data sets.
Settings of our computations are given in Section~\ref{sec-5-1}.
For the sake of clarity, we only present the results of $\SNR= 10$.
Those of $\SNR= 5, 15$ and $20$ display similar
phenomenon and are omitted.

The FA threshold ($>$0.3003) and Smooth-FA threshold ($>$0.2803) are
tentatively chosen such that their identified results share the same
sensitivity as $\chiK$-$\FDR_L$. The results of sensitivity and
specificity by all four methods are displayed in Table~\ref{table-3}.
Clearly, $\chiK$-$\FDR_L$ maximizes both the sensitivity ($0.8845$) and
the specificity ($0.9982$) in this example. $\chiK$-$\FDR$ achieves
similar specificity ($0.9957$) but slightly smaller sensitivity
($0.7522$) compared to $\chiK$-$\FDR_L$. However, in order to yield
comparable sensitivity ($0.8845$) as $\chiK$-$\FDR_L$, FA-threshold and
Smooth-FA-threshold approaches produce much lower specificities
($0.4012$ for FA; $0.6242$ for Smooth-FA).

%
\begin{table}
\tablewidth=270pt
\caption{Sensitivity and specificity, SNR${} = {}$10, the FDR control
level is 0.01} \label{table-3}
\begin{tabular*}{270pt}{@{\extracolsep{\fill}}lcc@{}}
\hline
& \textbf{Sensitivity} & \multicolumn{1}{c@{}}{\textbf{Specificity}}
\\
\hline
$\FA>0.3003$ & 0.8849 & 0.4012 \\
Smooth-FA $>0.2803$ & 0.8852 & 0.6242 \\
$\chiK$-FDR & 0.7522 & 0.9957 \\
$\chiK$-$\FDR_L$ & 0.8845 & 0.9982 \\
\hline
\end{tabular*}
\end{table}

\begin{figure}

\includegraphics{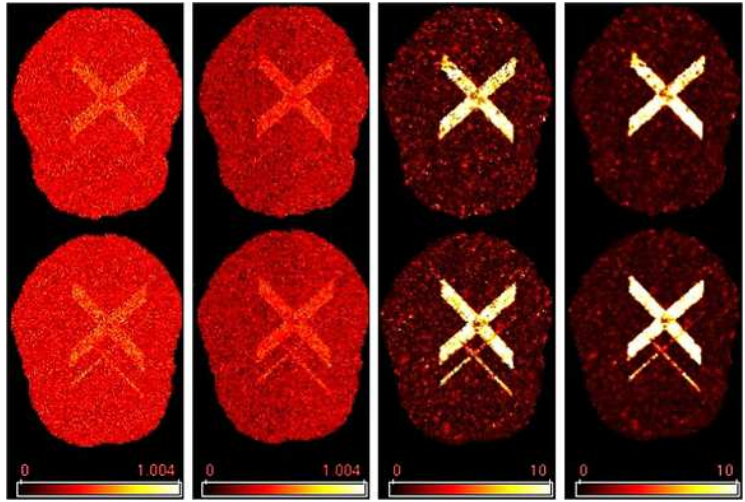}

\caption{Comparison of color maps for simulated data set. From left to
right: FA; Smooth-FA; $-\log(p)$
by $\chiK$; $-\log(\tilde p)$ by $\chiK$. $\SNR= 10$.} \label{Figure-6}
\end{figure}

\begin{figure}[b]

\includegraphics{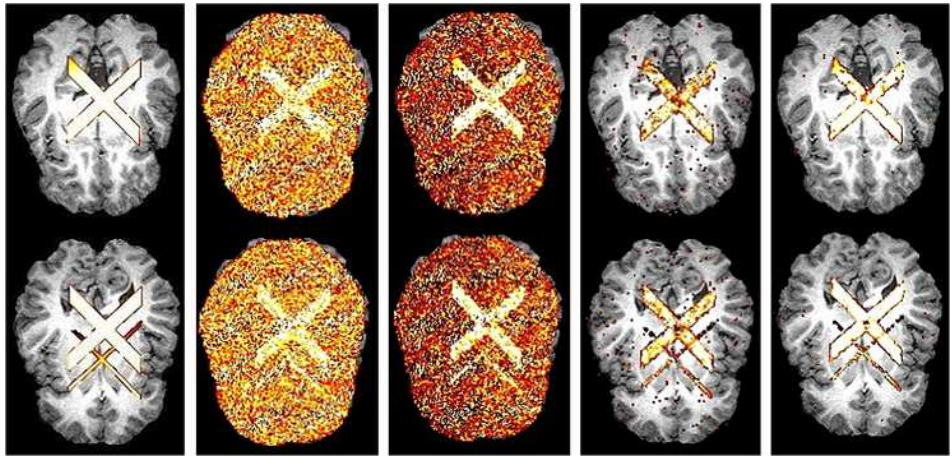}

\caption{Comparison of brain anisotropic areas discovered
for the simulated data set. From left to right: simulated brain
anisotropic areas;
$\FA>0.3003$; $\mathrm{Smooth}\mbox{-}\FA>0.2803$; $\chiK$-$\FDR$; $\chiK$-$\FDR_L$.
$\SNR= 10$. The control level is
$0.01$.}\label{Figure-7}
\end{figure}

The performances of the four methods are further compared on two
selected axial slices. Throughout our simulation and
real data examples, we apply the same registration transformations
from the brain data to the T1 high-resolution image of the subject's
brain. The two slices with the
simulated brain anisotropic areas highlighted are given in the leftmost
panel of Figure
\ref{Figure-7}. Figure~\ref{Figure-6} displays the color maps of the
FA, Smooth-FA, $-\log(p)$ and $-\log(\tilde p)$, with all $-\log(p)$
and $-\log(\tilde p)$ values greater than 10 set equal to 10 to improve
the visualization. Figure
\ref{Figure-7} compares the detected anisotropic areas by $\FA>0.3003$,
Smooth-$\FA>0.2803$,
$\chiK$-FDR and $\chiK$-$\FDR_L$ for $\SNR
=10$.

%

%
%

As clearly evidenced in Figures~\ref{Figure-6} and~\ref{Figure-7},
$\chiK$-FDR and $\chiK$-$\FDR_L$ not only provide detected results with
better accuracy than FA-threshold and Smooth-FA-threshold approaches,
but also yield better contrasts between the significant areas and the
nonsignificant ones. In contrast, the results from FA ($>$0.3003) and
Smooth-$\FA$ ($>$0.2803) not only fail to detect some truly significant
voxels, but also present tiny scattered faulty findings, which expect
to contaminate the downstream fiber tracking results. It has been seen
in the second to right and rightmost panels of Figures~\ref{Figure-6}
and~\ref{Figure-7} that the detected results by $\chiK$-FDR and
$\chiK
$-$\FDR_L$ well capture the primary features of the simulated
anisotropic areas. Compared with $\chiK$-FDR, the $\chiK$-$\FDR_L$
approach offers slightly more accurate identifications in both
isotropic and anisotropic water diffusion areas.

\section{Real data example} \label{sec-6}

We apply our proposed testing procedures on five subjects, whose DWIs
were acquired by the magnetic resonance (MR) experiments described below.

The brain magnetic resonance images (MRIs, including DWI and fMRI) of
each subject were
acquired with a GE SIGNA 3-T scanner equipped with high-speed
gradients and a whole-head transmit-receive quadrature birdcage
headcoil (GE Medical Systems). The anatomical scan for each subject
took approximately 20 minutes
[\citet{Dalton2005}]. In the anatomical scanning, the size of each voxel
in an $xy$-plane is
$0.9375\mbox{ mm}\times0.9375\mbox{ mm}$, field of view${} =
24\mbox{ cm}^2$, matrix${} = 256\times256$; $30$ axial slices are
acquired along the $z$-axis, slice
thickness${} = 3$~mm. A single reference image at $b=0$ and 12
diffusion-attenuated images with noncollinear directions of diffusion
gradients at $b=1000\mbox{ s}/\mbox{mm}^2$ were obtained. Since we focus
on the analysis of the anatomical structures
of the human brain in this paper, the detailed information for the functional
scans is omitted.

%
%
\begin{figure}[b]
\vspace*{-3pt}
\includegraphics{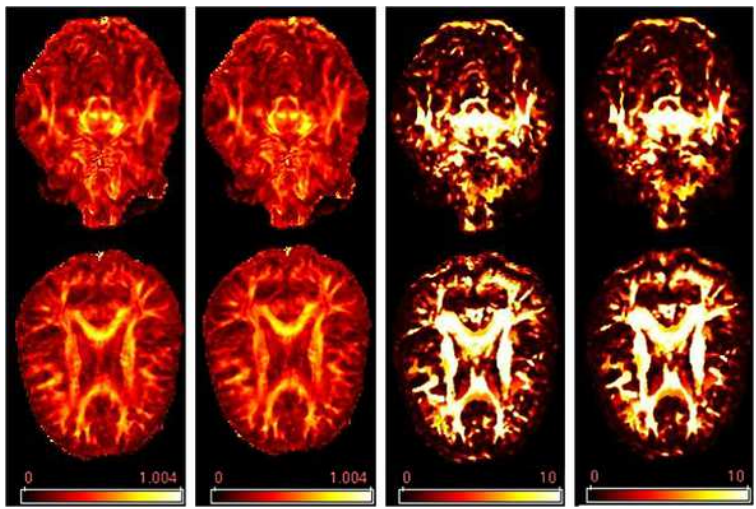}

\caption{Comparison of color maps for the real brain data
set. From left to right: FA; Smooth-FA; $-\log(p)$ by $\chiK$;
$-\log (\tilde p)$ by $\chiK$.} \label{Figure-8}
\end{figure}

Using the DWI data of a single subject as the representative, we first
present and compare the results by all four methods on two selected
axial slices of the brain. The results for the other four subjects
display similar scenarios and are omitted.

The acquired data set contains $256\times
256 \times30 = 1\mbox{,}966\mbox{,}080$ voxels with $400\mbox{,}309$ voxels located inside the brain.
In each voxel, the DT is estimated from regression model (\ref{eq-21}).
After that, the corresponding eigenvalues are obtained by
Schur decomposition.

%
\begin{figure}

\includegraphics{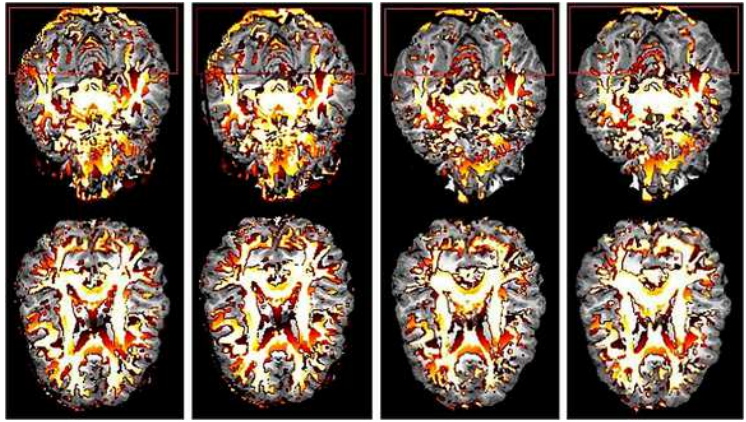}

\caption{Comparison of brain anisotropic areas discovered
for the real brain data set. From left to right: $\FA>0.35$;
$\mathrm{Smooth}\mbox{-}\FA>0.35$; $-\log(p)$ by $\chiK$; $-\log(\tilde p)$ by
$\chiK$.
The control level is 0.01.}\label{Figure-9}
\end{figure}

%
\begin{figure}[b]

\includegraphics{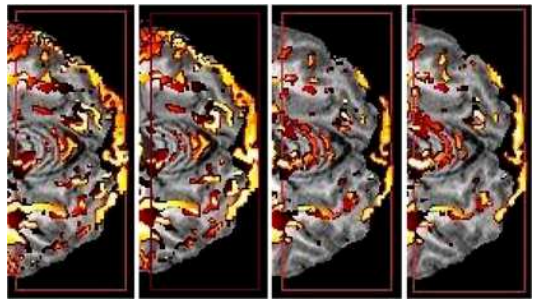}

\caption{Enlarged areas in the rectangles of Figure \protect\ref
{Figure-9}. Rotated $90^\circ$.} \label{Figure-10}\vspace*{-3pt}
\end{figure}

All settings are given in Section~\ref{sec-5-1}. The color maps of FA,
Smooth-FA, $-\log(p)$ and
$-\log(\tilde p)$ are displayed in Figure~\ref{Figure-8} on two
selected axial slices, whereas the corresponding detected anisotropic
diffusion brain areas by all four methods are provided in Figure \ref
{Figure-9}.
As evidenced in Figure~\ref{Figure-9}, compared with the identified
anisotropic areas by $\chiK$-FDR or $\chiK$-$\FDR_L$, FA-threshold and
Smooth-FA-threshold approaches produce more noisy detections. For
example, inspection of areas highlighted by red rectangles in the top
panels of Figure~\ref{Figure-9} (enlarged in Figure~\ref{Figure-10}),
FA $>0.35$ and Smooth-FA $>0.35$ detect more scattered tiny areas than
$\chiK$-FDR and $\chiK$-$\FDR_L$. Those are highly likely to be faulty
findings. Furthermore, an overview of the areas located close to the
top of the highlighted areas, FA $>0.35$ and Smooth-FA $>0.35$ present
more findings than $\chiK$-FDR and $\chiK$-$\FDR_L$. However, those
areas are located close to the boundary of the brain, and are most
likely to be nonfiber areas. In the meantime, as illustrated in Section
\ref{sec-2-2}, since the number of gradients $r$ in each voxel is
small, FA-threshold and Smooth-FA-threshold approaches cannot clearly
infer and control the error rate of the identification, and therefore
lack the rigorous criterion in selecting the appropriate thresholds in practice.
The superiority of our proposed methods to FA-threshold and
Smooth-FA-threshold approaches can be further illustrated by Figure
\ref{Figure-8}, the color maps of FA, Smooth-FA, $-\log(p)$ and
$-\log(\tilde p)$. The color maps of both $-\log(p)$ and
$-\log(\tilde p)$ show better contrasts between the
anisotropic and isotropic areas than those of FA and Smooth-FA,
indicating that $\chiK$-FDR and $\chiK$-$\FDR_L$
more effectively separate anisotropic diffusion areas from isotropic ones.

To further illustrate the efficacy of our methods in reducing the
scattered faulty findings, we summarize the frequency of identified
isolated anisotropic voxels over each subject's brain for all four
methods in Table~\ref{Table-freq}. In particular, denote by $\mathN(v)
= \{v' = (v_x', v_y', v_z')\dvtx  |v_i'-v_i| \le1, \mbox{ for all } i = x,
y, \mbox{ and } z \}$ the nearest neighboring voxels of $v$. Table
\ref
{Table-freq} displays the number of voxels carried in $\mathS_{1,m}$
(left) and $\mathS_{2,m}$ (right) over each subject's brain.
Here $\mathS_{u,m} = \{v\dvtx  \mbox{ method }m\mbox{ identify } u \mbox{
voxels in }\mathN(v) \mbox{ as anisotropic}\}$. Clearly, $\mathS_{1,m}$
carries voxels where only the voxel $v$ itself is identified by method
$m$ over voxels in $\mathN(v)$. Likewise, $\mathS_{2,m}$ contains
voxels where only two voxels, that is, the voxel $v$ itself and another
voxel, are identified by method $m$ over voxels in $\mathN(v)$. We
observe that voxels in $\mathS_{1,m}$ and $\mathS_{2,m}$ are highly
likely to be faulty findings by method $m$, since they are ``isolated''
from other identified anisotropic voxels; fiber tracts, in contrast,
are typically spatially connected.

%
\begin{table}
\caption{Number of voxels carried in $\mathS_{1,m}$ (left) and
$\mathS_{2,m}$ (right) over each subject's brain} \label{Table-freq}
\begin{tabular*}{\textwidth}{@{\extracolsep{\fill}}lcccccccccc@{}}
\hline
& \multicolumn{5}{c}{\textbf{Subject}} &\multicolumn
{5}{c@{}}{\textbf
{Subject}}\\[-4pt]
& \multicolumn{5}{c}{\hrulefill} &\multicolumn{5}{c@{}}{\hrulefill
}\\
\textbf{Methods} & \textbf{1} & \textbf{2} & \textbf{3} & \textbf
{4} &
\textbf{5}
& \textbf{1} & \textbf{2} & \textbf{3} & \textbf{4} & \textbf{5}
\\\hline
FA $>0.35$ & 323 & 360 & 348 & 356 & 280 &718 & 770 & 720 & 758 & 671\\
Smooth-FA${}>0.35$ & 279 & 288 & 327 & 298 & 227&775 & 793 & 657 & 721
& 660 \\
$\chiK$-FDR & 155 & 177 & 166 & 173 & 189&315 & 409 & 357 & 403 & 355
\\
$\chiK$-$\FDR_L$ & \phantom{0}63 & \phantom{0}77 & 100 & \phantom{0}93
& 103& 129 & 158 & 149 & 212 & 138\\
\hline
\end{tabular*}
\end{table}

It has been seen from Table~\ref{Table-freq} that our methods continue
to outperform FA-threshold and Smooth-FA-threshold approaches by
producing much less isolated findings. Compared with $\chiK$-FDR,
$\chiK
$-$\FDR_L$ produces even a smaller number of isolated identifications.
Such a result is not surprising considering what has been observed in
Figures~\ref{Figure-6}--\ref{Figure-9}. Combining all the numerical
results above, we therefore recommend the identifications by $\chiK
$-$\FDR_L$ as the final results.

\section{Discussion} \label{sec-7}

In DTI studies, one of the important research topics is to refine the
identification of the anisotropic water diffusion areas of human brain
in vivo. There are two general strategies aiming to address this
problem. The\vadjust{\goodbreak} first is to improve the DT estimation. A downstream
procedure is then needed to identify the anisotropic water diffusion
areas. The second is to refine the construction of the scalar
measurements or establish more powerful test statistics for every brain
voxel. The identification is then based on thresholding the
measurements or a certain testing procedure. We observe that the second
provides more intrinsic insight into the water diffusivity in each
voxel, and therefore is more effective in allusion to the
identification of anisotropic water diffusion areas.

From an experimental point of view, there are two ways to improve the
acquisition schemes for the DWI data. One is to increase the number of
diffusion gradients for every brain voxel, the other is to improve the
resolution of the imaging space, that is, increase the number of brain
voxels. To the best of our knowledge, existing methods for constructing
scalar measurements or test statistics are all single voxel based.
Therefore, the corresponding inferences improve only when the number of
diffusion gradients in each voxel increases, while ignoring the
possible improvement of the resolution of the imaging space. The
methods proposed in this paper fill this gap by incorporating the
eigenvalues in the neighboring voxels in the construction of the test
statistic $\chiK$.

In this study we have established the asymptotic distribution of our
proposed test statistic. One of the main assumptions required by our
theoretical results is that the number of neighboring voxels for
constructing $\chiK$ is large. This assumption can be well achieved
when the resolution of the imaging data is high. As such, the bias
components carried in the eigenvalue estimates no longer play a key
role in the identification of anisotropic water diffusion brain areas.
In both simulation and real data analysis, we have observed that our
proposed $\chiK$-FDR and $\chiK$-$\FDR_L$ approaches lead to different
identification results from FA-threshold and Smooth-FA-threshold
approaches, popularly adopted in the DTI community. In particular, the
scattered findings by our methods are much less than those by
FA-threshold and Smooth-FA-threshold approaches, indicating that by
incorporating neighboring information, our methods are capable of
screening out those isolated voxels which are highly likely to be
faulty findings. Results based on simulated DWI data demonstrate that
our proposed test statistic $\chiK$ agrees reasonably well with the
$\chi^2$ distribution when $n=25$ (or larger), and our methods achieve
better accuracy than FA-threshold and Smooth-FA-threshold approaches in
the identification of anisotropic brain voxels. Furthermore, the
Smooth-FA-threshold approach is capable of partially solving the bias
problem in the eigenvalue estimates [\citet{PolzehlTabelow2009}].
However, the performance of the approach heavily replies on the
estimation of the heteroscedastic
variances over the entire brain. These variances, in turn, are modeled
by a linear model and estimated using the reference signals $\phi_0(v)$.
We observe based on simulation studies that when the reference
signals over the entire brain are not homogeneous or do not share
comparable variances as attenuated signals $\phi_i(v)$, the performance
of the approach varies\vadjust{\goodbreak} [see more simulation results provided in the
supplemental document: \citet{Yu2012}]. In contrast, under all these
cases, our proposed $\chiK$-$\FDR_L$ approach consistently offers
descent results. We therefore conclude that over all four methods, our
proposed $\chiK$-$\FDR_L$ approach performs the best over all our
simulation studies. Unlike the simulation examples, we are unable to
show the true anisotropic (fiber) areas of the human brains in real DTI data.

We would like to point out that in DTI studies, identification of
anisotropic water diffusion areas is just one step of the full
analysis. Downstream analysis, such as fiber tracking, is usually
needed to fully capture the physical structure of the human brain.

In this paper we have focused on establishing testing procedures to
distinguish anisotropic DT voxels from isotropic ones based on second
order DT models. Our proposed methods have been compared with the
FA-threshold and Smooth-FA-threshold approaches. The results of this
paper can serve as a benchmark in analyzing the anatomical structures
of human brains based on DTI data. We observe that the following topics
are highly related to this paper, and may absorb the interest of the
community in future research:
\begin{itemize}
\item With the idea of incorporating spatial information,
single-voxel-based approaches, including FA, can possibly be improved
by appropriately accounting for the information in the neighboring
voxels. Furthermore, in this paper, $\chiK$ is established on the DTs
estimated from model (\ref{eq-21}). Similar approaches can be
constructed based on more sophisticated DT estimates from other approaches.
\item A similar strategy as that in this paper can be adopted to
establish testing procedures for teasing apart the morphologies of
anisotropic DTs. Furthermore, the basic ideas can be further extended
to establish testing procedures for identifying the presence of signals
for data with spatial structures, though we focus on DTI data in this paper.
\item Another popular topic in DTI research is to consider higher-order
tensor models [\citet{Grigis2011} and therein], which are powerful tools
for investigating the fiber structures when there are several fiber
bundles in a single voxel. The local test idea in this paper can be
possibly extended to identify the number of intersected fiber bundles
based on those higher-order tensor models.\vspace*{-1pt}

\end{itemize}

\section*{Acknowledgments}
We thank the Editor, the Associate
Editor and four referees for invaluable suggestions, which greatly
improved the quality of this paper.\vspace*{-1pt}

\begin{supplement}[id=suppA]
\stitle{Supplement to ``Local tests for identifying anisotropic
diffusion areas in human brain with DTI''}
\slink[doi]{10.1214/12-AOAS573SUPP}  
\sdatatype{.pdf}
\sfilename{aoas573\_supp.pdf}
\sdescription{This file provides proofs for Theorems 1 and 2, a short
summary of\vadjust{\goodbreak}
$\FDR_L$ procedure [Zhang, Fan and Yu (\citeyear{Zhang2011})], steps for constructing
$\chiK$ based on DWI data and some more simulation results.}
\end{supplement}

%


\printaddresses

\end{document}